# The Geometrically Frustrated Spin Glass (Fe$_{1-p}$Ga$_p$)$_2$TiO$_5$

Y. Li[1], D. Phelan[1], F. Ye[2], H. Zheng[1], E. Krivyakina[1,3], A. Samarakoon[1], P. G. LaBarre[4], J. Neu[5,6], T. Siegrist[5,7], S. Rosenkranz[1], S. V. Syzranov[4], A. P. Ramirez[4]

[1]*Materials Science Division, Argonne National Laboratory, Lemont, IL, 60439*
[2]*Neutron Scattering Division, Oak Ridge National Laboratory, Oak Ridge, TN, 37830*
[3]*Department of Physics, Northern Illinois University, DeKalb, Illinois 60115, USA*
[4]*Physics Department, University of California Santa Cruz, Santa Cruz, CA 95064*
[5]*NHMFL, Florida State University, Tallahassee, FL 32310*
[6]*Nuclear Nonproliferation Division, Oak Ridge National Laboratory, Oak Ridge, TN, 37831*
[7]*Department of Chemical and Biomedical Engineering, FAMU-FSU College of Engineering, Tallahassee, FL, 32310*

## ABSTRACT

The unusual anisotropy of the spin glass transition in the pseudobrookite system Fe$_2$TiO$_5$ has been interpreted as arising from an induced, van der Waals-like, interaction among magnetic clusters. Here we present susceptibility ($\chi$) and specific heat data ($C$) for Fe$_2$TiO$_5$ diluted with non-magnetic Ga, (Fe$_{1-p}$Ga$_p$)$_2$TiO$_5$, for disorder parameter $p = 0$, 0.11, and 0.42, and elastic neutron scattering data for $p = 0.20$. A uniform suppression of $T_g$ is observed upon increasing $p$, along with a value of $\chi(T_g)$ that increases as $T_g$ decreases, i.e. $d\chi(T_g)/dT_g < 0$. We also observe $C(T) \propto T^2$ in the low temperature limit. The observed behavior places (Fe$_{1-p}$Ga$_p$)$_2$TiO$_5$ in the category of a strongly geometrically frustrated spin glass.

The pseudobrookite compound $Fe_2TiO_5$ is unusual for exhibiting spin glass (SG) freezing signatures – a cusp in susceptibility, $\chi(T)$, at the glass transition temperature, $T_g$, and hysteresis in $\chi(T)$ below $T_g$ – only with the applied magnetic field along the crystallographic ***c***-axis [1]. When probed in the ***a*** and ***b*** directions, no anomaly is seen at $T_g$, which is highly unusual because the magnetic ion, $Fe^{3+}$, is an *s*-state ion and thus possesses no single-ion anisotropy. In addition, the Fe and Ti atoms randomly occupy the two cation sites. When a similar *s*-state ion, $Mn^{2+}$, resides on a periodic lattice, such as in $MnF_2$, anisotropy develops below the Neel transition due to the dipole-dipole energy, which establishes an ordering direction. It seems unlikely, therefore, given the randomness of the $Fe^{3+}$ ions, that this mechanism can explain the anisotropy in $Fe_2TiO_5$. Thus, it is also likely impossible to describe the complete anisotropy in its spin glass response using atomic spins as the freezing degree of freedom. Recently we revisited this problem and showed, using neutron scattering, that the Fe spins develop nano-sized regions of antiferromagnetic (AF) order where the spins are either aligned or anti-aligned with the long direction, parallel to the ***a***-axis [2]. These regions take the shape of surfboards with correlations lengths of 40 Å × 10 Å along the ***a***- and ***c***-axis, respectively, but along the ***b***-axis, only nearest neighbor spins are correlated due to frustration. We also showed that the interaction between the surfboards comes from the fluctuations of their magnetization along the ***c***-direction, i.e. the direction transverse to the ordered spins. Such a *magnetic van der Waals* force can thus explain, at least qualitatively, the anisotropy of freezing in $Fe_2TiO_5$.

An important aspect of $Fe_2TiO_5$ is its strong geometrical frustration, as indicated by a large frustration parameter, $f \equiv \theta_W/T_g = 900/55 \approx 16$ [3], where $\theta_W$ is the Curie-Weiss temperature. As described in recent work by two of us [4], when varying the amount of quenched disorder in strongly geometrically frustrated (GF) systems, the SG transition temperature $T_g$ and magnetic susceptibility $\chi(T_g)$ follow the universal trend $d\chi(T_g)/dT_g < 0$, which is opposite to the trend $d\chi(T_g)/dT_g > 0$ displayed by conventional SGs. Furthermore, $T_g$ in strongly GF materials grows with decreasing the density of vacancies [4], the dominant source of quenched disorder. This trend calls into question the achievability of quantum spin liquids widely sought in these materials, because it suggests that the SG state, believed to be incompatible with quantum spin liquids, occurs even in very clean samples.

Although $Fe_2TiO_5$ belongs to the same class of strongly GF materials, it is distinct from other GF materials in that it develops the described surfboard-shaped and these clusters apparently play an essential role in the SG freezing. The strong interaction between these clusters may be responsible, for example, for higher SG-transition temperatures in $Fe_2TiO_5$ than in the other GF systems [4]. The SG transition and the structure of the glass state in $Fe_2TiO_5$ thus require a separate careful investigation.

*Summary of results.* Here we present $\chi(T)$ and specific heat ($C(T)$) results for the dilution series $(Fe_{1-p}Ga_p)_2TiO_5$ for $0 \leq p \leq 0.42$, in which non-magnetic $Ga^{3+}$ substitutes for $Fe^{3+}$, as well as neutron scattering data for $p = 0.20$. Similar to other GF systems where $f \gtrsim 10$, we find that $T_g$ decreases with increasing susceptibility, $d\chi(T_g)/dT_g < 0$, as disorder is increased with Ga content, $p$. The specific heat also displays a low-temperature quadratic-in-$T$ term that grows with $p$, similar to that seen in other GF materials. Our neutron scattering measurements indicate the existence of surfboard-shaped correlated regions, like in $Fe_2TiO_5$, but substantially shrinking, as $p$ is increased. Thus, whereas $T_g$ reduction in other GF systems is due to an increase in vacancies around which "quasispins" form and undergo freezing, in $(Fe_{1-p}Ga_p)_2TiO_5$ the quasispins are AF-ordered surfboard-shaped regions. This distinction may explain why the energy scale for SG freezing is larger in the present material than in other GF systems. Nevertheless, in both types of systems the quasispin is a composite object, i.e. formed from many atomic spins and, thus noted, the argument of Ref. [4] pertains.

*Synthesis and characterization.* In the present study, we used crystals of $(Fe_{1-p}Ga_p)_2TiO_5$, grown by J. P. Remeika [5] for measurements of magnetic susceptibility and specific heat. Because these crystals were grown from a flux, their nominal concentrations serve only as a coarse guide to the resultant concentrations. Also, since techniques for atom identification in the solid state, such as energy dispersive atomic spectroscopy, possess systematic uncertainty in ratios between constituent atoms, we rely here on the susceptibility itself to determine the concentration of Fe in each sample. This approach relies on our previously reported measurement of $\chi(T)$ in $Fe_2TiO_5$ for temperatures between 400 K and 900 K, which yielded an effective moment of $\mu_{eff} = 6.12 \pm 0.05\ \mu_B$, in reasonable agreement with the value $\mu_{eff} = 5.92\ \mu_B$ expected for a free S = 5/2 ion [2]. Since the most concentrated compound of the series yielded an effective moment close to the theoretical value, it is reasonable to expect that the effective moment in more dilute compounds will be identical since the chemical environment and oxidation states are the same. We note that, since $Fe^{3+}$ and $Ti^{4+}$ can each occupy either the A or B sites in the pseudobrookite $A_2BO_5$ structure, even for $p = 0$, the Fe occupation is random. In addition, due to the different oxidation states of Fe and Ti, the pure pseudobrookite compound, $Fe_3O_5$, cannot be made with only trivalent Fe ions, i.e. with zero magnetic disorder. Since $Ga^{3+}$ is isovalent with $Fe^{3+}$ but non-magnetic and similar in size, substituting it for $Fe^{3+}$ will increase magnetic disorder.

*Measurements.* The crystal structure of the single crystals used here was confirmed by X-ray diffraction to be the pseudobrookite phase for $p = 0$, 0.11 and 0.42. The orthorhombic lattice constants $\{a, b, c\}$ in Ångstroms for $p = 0$, 0.11, and 0.42 are {3.732(1), 9.8125(2), 10.0744(2)}, {3.7023(1), 9.7872(2), 9.9670(2)} and {3.6857(1), 9.7903(2), 9.9681(2)} respectively. The dependence of $\{a, b, c\}$ on $p$ is shown in the upper left inset of Fig 1., normalized to their $p = 0$ values. Magnetization measurements were performed in two different Quantum Design Magnetic Property Measurement Systems (MPMS3s). For $\chi(T < 300 \text{ K})$, a conventional sample holder was used. For $\chi(T > 300 \text{ K})$, the sample was mounted with Zircar cement on a rod designed for high-temperature measurements. The specific heat, $C(T)$, was measured for $T > 2.5K$ using the relaxation technique in a Quantum Design Physical Property Measurement System (PPMS). For $T < 4K$, $C(T)$ was measured using the semi-adiabatic technique in a top-loading dilution refrigerator. Energy dispersive X-ray absorption spectroscopy (EDAXS) was performed on the samples and yielded a Fe:Ga ratio that confirmed the general trend expected for Ga substitution. Neutron diffraction measurements were performed on a crystal of nominal composition $Fe_{1.6}Ga_{0.4}TiO_5$ ($p = 0.20$) which was grown in a floating zone furnace at Argonne National Laboratory. The neutron scattering measurements were performed on the time-of-flight instrument Corelli (ORNL) using a pseudo-white beam [6]. This is the same instrument which was used to measure $Fe_2TiO_5$, as reported previously [2].

*Magnetic susceptibility.* In Figure 1 are shown $\chi(T)$ data for $(Fe_{(1-p)}Ga_p)_2TiO_5$ with $p = 0$, 0.11, and 0.42. It is important to recognize that, since Fe, Ga, and Ti can each occupy either of the two cation sites, the magnetic ion stoichiometry $p = 0$, 0.11, and 0.42 corresponds to Fe occupancies of $n_{Fe} = 0.67$, 0.59, and 0.39. We see that $T_g$ decreases with decreasing concentration of magnetic ions – $T_g$ = 55 K, 27 K, and 18 K, for $p = 0$, 0.11, and 0.42 respectively. While a decrease in $T_g$ with decreasing concentration of magnetic ions is also seen in conventional SG systems such as *Cu*Mn [7], *Au*Fe [8, 9], and $Eu_xSr_{1-x}S$ [10], the increase of $\chi(T_g)$ with decreasing $T_g$, $d\chi(T_g)/dT_g < 0$, is not presently understood, as pointed out in Ref. [4]. As discussed in [4], such behavior is, however, a distinct signature of the spin glass transition in all other GF for which $f \geq 10$. The inset of Figure 1 shows the inverse susceptibility for the same compounds where, as previously indicated, the $p = 0.11$ and 0.42 data have been scaled to exhibit the same slope and thus the same $\mu_{eff}$ as the $p = 0$ data.

*Specific heat*. In Fig. 2 are shown specific heat data for $p = 0$, 0.11, and 0.42 over different temperature ranges. In Fig. 2 (a), $C(T)/T$ versus $T$ up to 300K is plotted per "mole-formula unit", appropriate for comparing the main contributions, namely lattice and magnetic excitations. These data show a broad hump centered at approximately 140 K, the high-temperature decrease of which indicates an approach on warming into the Dulong-Petit region. The data in this frame are mainly due to lattice excitations, as can be judged from the area of the shaded region which is equal to the total magnetic entropy for $p = 0$ and approximately 10% of the total entropy from T = 0 to 300K. As the Fe concentration decreases, one can see the height of the broad peak decrease slightly while the entropy below 50K increases. This entropy shift may be indicating that, as Fe site-occupancy decreases, a greater fraction of spins find themselves in the low-energy region of the rough energy landscape associated with random freezing. In Fig. 2 (b) are shown data at intermediate temperatures. Here, the differences among the data sets reflect the entropy transfer mentioned above. Finally, in Fig. 2 (c) are shown $C(T)/T$ for $T$ down to 200 mK, below which an additional, possibly nuclear, contribution appears as a rise in $C(T)/T$ on decreasing $T$ for the last few points, especially noticeable for $p = 0.42$. Taking this contribution into account, the data are consistent with a linear term that is too small to be unambiguously identified. More apparent is a temperature dependence below $T = 3K$ that is consistent with $C(T) \propto T^2$, behavior which has been seen in several other strongly GF systems [11-14]. While the precise origin of the $C(T) \propto T^2$ behavior is not well understood [15], it is reasonable to ascribe the associated excitations to inter-surfboard spins, and not to intra-surfboard spin waves, since these are expected to follow $C(T) \propto T^3$, albeit with a strong finite-size effect, which should produce a gap in the spin wave spectrum. In general, the specific heat of a disordered spin system may come from itinerant spin modes, such as Halperin-Saslow [15, 16] modes or spin-lattice "photons" [17-21], or from localized excitations [16]. Such itinerant modes can result in the $C(T) \propto T^2$ in an effectively 2D, e.g. layered, system. In 3D, however, they will lead to the $C(T) \propto T^3$ temperature dependence, inconsistent with our observations here.

It is reasonable, therefore, to ascribe the observed quadratic specific heat to localized excitations among inter-surfboard spins. The quadratic dependence of the specific heat requires that the density of states (DoS) of these excitations (the density of levels per unit energy in an ensemble of excitations), $\nu(E)$, be linear in energy $E$. As shown in Ref. [22], in a strongly disordered medium with short-range interactions, a linear-in-$E$ DoS is a universal feature of excitations corresponding to shifting any conserved quantity between nearby sites (quasi-localized states). The role of such a quantity may be played by spin density or "charge" in the Coulomb phase [17, 18, 23-25] of a GF magnet. The corresponding mechanism requires that the shifts of the conserved quantity do not become arbitrarily large (which is possible in metastable states such as spin glasses). Because the DoS of excitations is field-independent in strongly disordered systems, the respective contribution to the specific heat will be field-independent, as is observed between $H = 0$ and 6 T for the $p = 0.42$ compound, shown in Fig. 2 (c). We leave, however, an investigation of the exact nature of the corresponding excitations for future studies.

*Neutron scattering*. Despite similarities with other GF systems, an important distinction can be drawn since, in $(Fe_{(1-p)}Ga_p)_2TiO_5$, the degrees of freedom that undergo SG freezing are postulated to be the surfboard-shaped ordered regions seen in neutron scattering [2]. The size of these regions in the undoped limit ($p = 0$) is approximately 40 Å × 3 Å × 10 Å, and thus an order of magnitude larger in the *a*-direction than the size of typical vacancy-induced quasi-spins which are thought to undergo freezing in other GF systems [26, 27]. Indeed, the large surfboard size, coupled with the observation that the ordered intra-surfboard spins point along the ***a***-axis, whereas spin freezing is seen only along the ***c***-axis, suggested a surfboard-surfboard interaction induced by fluctuations in the magnetization that are transverse to the spin ordering direction in the surfboards [2]. Below we describe the evolution of surfboard size and shape with $Fe^{3+}$ dilution.

Comparison of the magnetic diffuse neutron diffraction patterns of the $p = 0$ and $p = 0.20$ is shown in Fig. 3 and Fig. 4. Fig. 3a and 3b contrast the data measured in the (½ *k l*) plane for these two compositions. The peaks in scattering intensity are highly elliptical for $p = 0$, being broad along *k* and sharper along *l*, which reflects the differing correlation lengths along these two directions. Correlations along ***b*** are limited to nearest neighbors; however, along ***c*** the correlation length is about 11 Å as determined from the inverse of the half-width-half maximum of Lorentzian fits of a cut through the peaks along *l* as shown in Fig. 3c. These results are consistent with our previous results for $p = 0$ [2]. For $p = 0.20$, the peaks occur at the same positions in reciprocal space as for $p = 0$, indicating the continued presence of surfboard-shaped correlation regions for smaller Fe site occupancy. The peak widths along $k$ indicate that the correlations are similarly nearest neighbor-limited along **b**. However, the peaks are significantly broader along $l$ for $p = 0.20$, and the similarly performed fits in Fig. 3d indicate that the correlation length is significantly reduced at $T = 5$ K to only ~ 3 Å, approaching the correlation length along **b**. Further, it is quite apparent that the peaks in the ($h$ 0 $l$) planes (Fig. 4a and 4b) are broader for $p = 0.20$ than for $p = 0$. Fits along $h$ indicate that, despite considerable reduction induced by Ga dopants, the correlation length along **a** for $p = 0.20$ at 5 K remains significantly longer than the lattice constant, with a value of ~11 Å, resulting in regions with a size ~ 11 Å × 3 Å × 3 Å. Thus, while the aspect ratio of the correlated regions remains surfboard-like, the shape has become more prolate than for $p = 0$.

*Discussion*. The decrease in surfboard size with increasing $p$ is consistent with the associated decrease in $T_g$: smaller surfboard sizes correspond to smaller fluctuations of the magnetization on the surfboard, weaker interactions between the surfboards and, concomitantly, lower glass transition temperature $T_g$ (in accordance with the quantitative description developed in Ref. [2] for the $p = 0$ case). At the same time, the density of inter-surfboard spins is also decreasing as $p$ increases, which should also contribute to the observed decrease in $T_g$. An accurate quantitative description of the behavior of $T_g$ will require the determination of the inter-surfboard distance and other microscopic details whose investigation we leave for future studies.

We noted already that the behavior of the magnetic susceptibility and the glass transition temperature follow the trend $d\chi(T_g)/T_g < 0$ that has been shown [4] to persist universally in all geometrically frustrated magnets. This trend is accompanied by the growth of the glass-transition temperature with decreasing vacancy density that implies the existence of a ``hidden energy scale'' introduced recently in Ref. [4] that determines the transition temperature in the perceived vacancy-disorder-free limit. We note that this limit cannot be reached in (Fe$_{(1-p)}$Ga$_p$)$_2$TiO$_5$ due to the different valence states of Fe and Ti. Nevertheless, it is reasonable to discuss (Fe$_{(1-p)}$Ga$_p$)$_2$TiO$_5$ in the same framework as the other strongly GF systems since it also exhibits $d\chi(T_g)/dT_g < 0$.

One of the possible scenarios [4] for the hidden energy scale is the interplay of weak residual disorder with a sharp crossover in the permeability, $\mu(T)$, of the clean medium, that occurs as a result of the interplay of magnetic interactions and entropy. In most systems [4], the hidden energy scale is on the order of $T^* \sim 10K$ which suggests that interactions significantly weaker than the short-range exchange interactions between the spins play a role. In (Fe$_{(1-p)}$Ga$_p$)$_2$TiO$_5$, the role of such interactions could be played by the induced magnetic interactions between the surfboards. Additionally, we see that the surfboard size decreases with decreasing atomic spin density, which would also aid in reducing $T_g$. Further theoretical work is needed to identify the role of the possible interactions to the hidden energy scale for spin freezing.

In summary, we have measured $\chi(T)$, $C(T)$, and elastic neutron scattering in single crystals of (Fe$_{(1-p)}$Ga$_p$)$_2$TiO$_5$,, for $p > 0$. We find that $d\chi(T_g)/T_g < 0$ as the spin glass $T_g$ is made to decrease with increasing $p$, similar to behavior seen in other GF magnets. The model proposed to explain the unusual anisotropy of the freezing anomaly at $T_g$ for $p = 0$ is validated for $p = 0.20$, in which correlated regions, albeit smaller than seen for $p = 0$, are observed with a concomitant reduction in $T_g$.

Acknowledgments

We thank A. Henderson for assistance on materials related to but not included in this study. Work at UCSC (PL, AR) was supported by the U.S. Department of Energy grant DE-SC0017862. Work at Argonne National Laboratory (YL, DP, HZ, EK, AS, SR) included crystal growth; neutron scattering measurements, analysis, and interpretation; and high temperature susceptibility and was supported by the U.S. Department of Energy, Office of Science, Office of Basic Energy Sciences, Materials Sciences and Engineering Division. A portion of this research used resources at the Spallation Neutron Source, a DOE Office of Science User Facility operated by Oak Ridge National Laboratory.  The work at FSU (JN and TS) was supported by the National Science Foundation grant NSF DMR-1606952. Part of the work was carried out at the National High Magnetic Field Laboratory, which is supported by the National Science Foundation under grant NSF DMR-1644779, and the State of Florida.

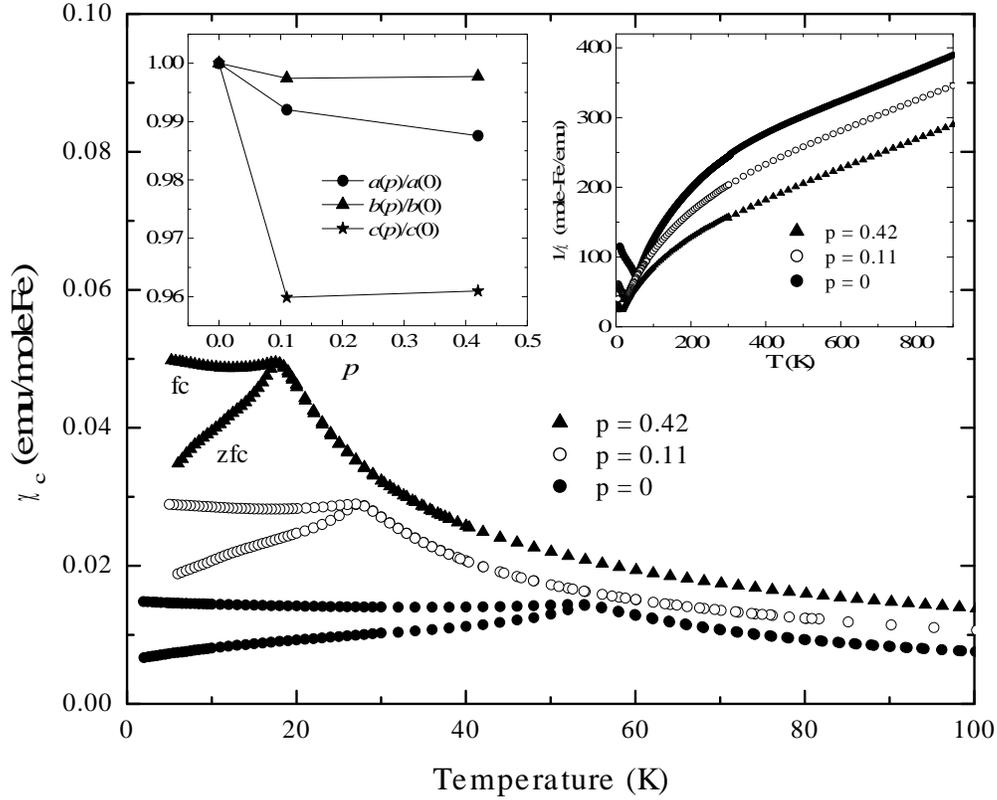

Figure 1. Susceptibility of $(Fe_{(1-p)}Ga_p)_2TiO_5$ for various values of $p$ measured with an applied field of H = 100 G. The lower curve at temperatures below each kink were taken after zero field cooling and the upper curves after field cooling. Inset: The inverse susceptibility for the same samples. The $p$-values for the samples other than $p = 0$ were determined by adjusting the molar mass to yield the same high-temperature slope of $1/\chi(T)$ as the $p = 0$ sample, the Fe concentration of which is known exactly.

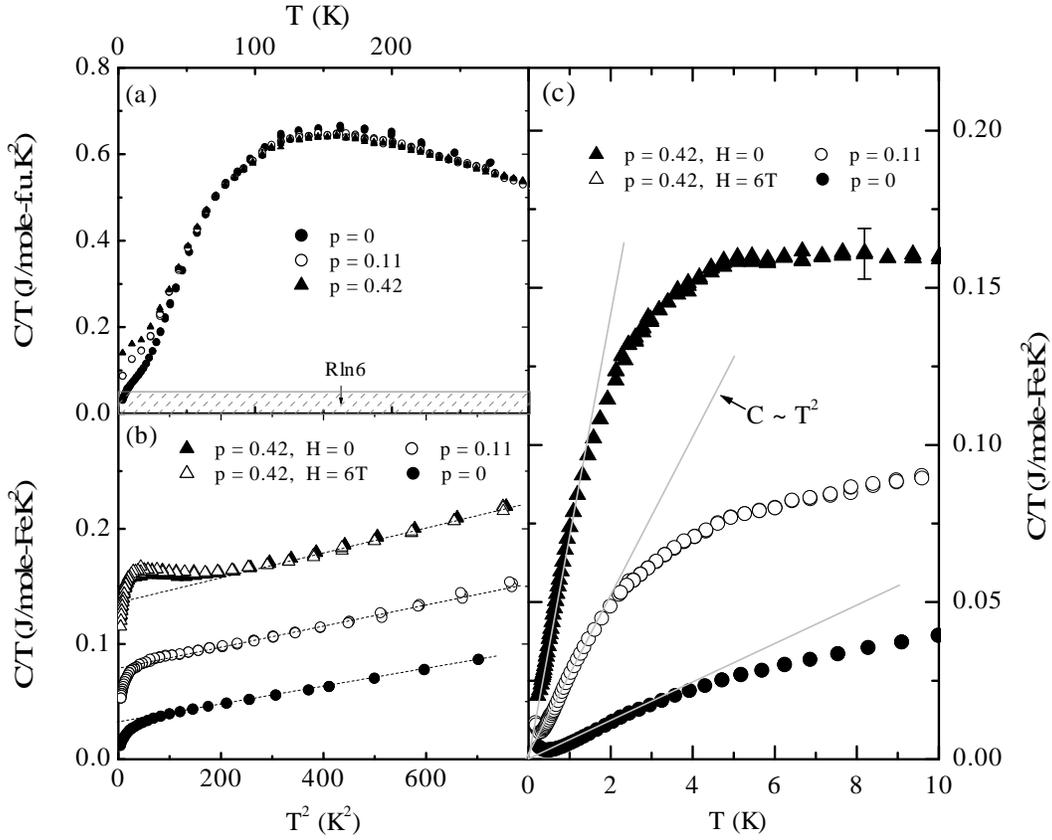

Figure 2. Specific heat divided by temperature of $(Fe_{(1-p)}Ga_p)_2TiO_5$ for various values of $p$. (a) $C(T)/T$ vs. temperature for $p = 0$ and 0.42, each showing two distinct short-range-order humps. (b) $C(T)/T$ vs $T^2$ in the intermediate temperature regime. (c) $C(T)/T$ vs $T$ for $T < 10$ K, showing the dominant $T^2$ behavior for $T < 3$K.

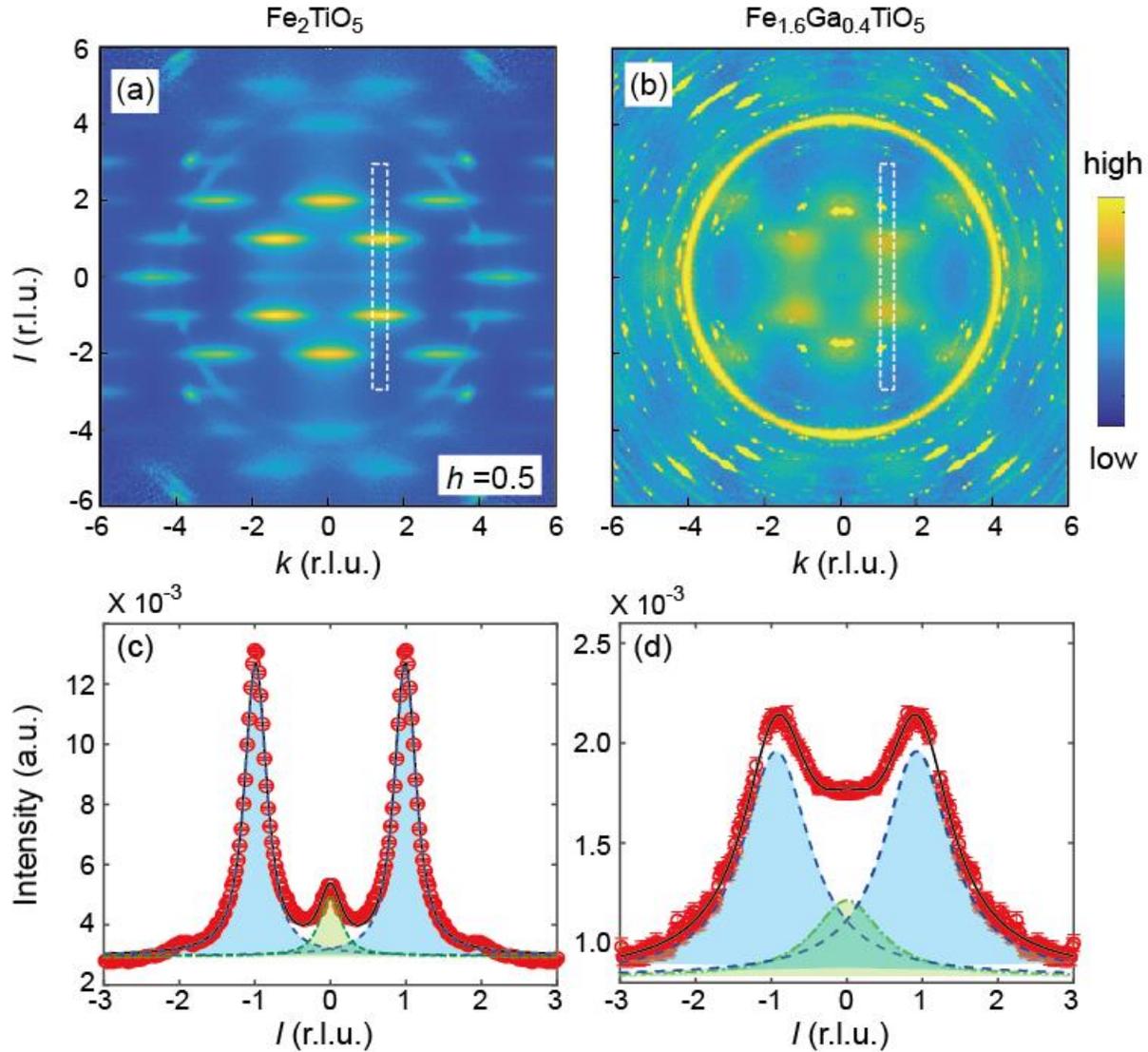

Figure 3. Total neutron scattering intensity in the [0.5, $k$, $l$] planes for $Fe_2TiO_5$ at T=5K (a) and $Fe_{1.6}Ga_{0.4}TiO_5$ ($p$=0.20) at T = 6K (b). (c),(d) Total neutron scattering intensity along the $l$ - direction as indicated by the dashed rectangle in (a) and (b), respectively. The intensity was fitted with three Lorentz functions convoluted with a Gaussian resolution function. The three Lorentz peaks are set to have the same HWHM whose reversal is the correlation length along the $l$ - direction and is estimated to be 3.0±0.1 Å at 6 K and 2.7±0.1 Å at 50 K for $Fe_{1.6}Ga_{0.4}TiO_5$ and 10.9±0.15 Å at 5 K and 9.6±0.15 Å at 60 K for $Fe_2TiO_5$

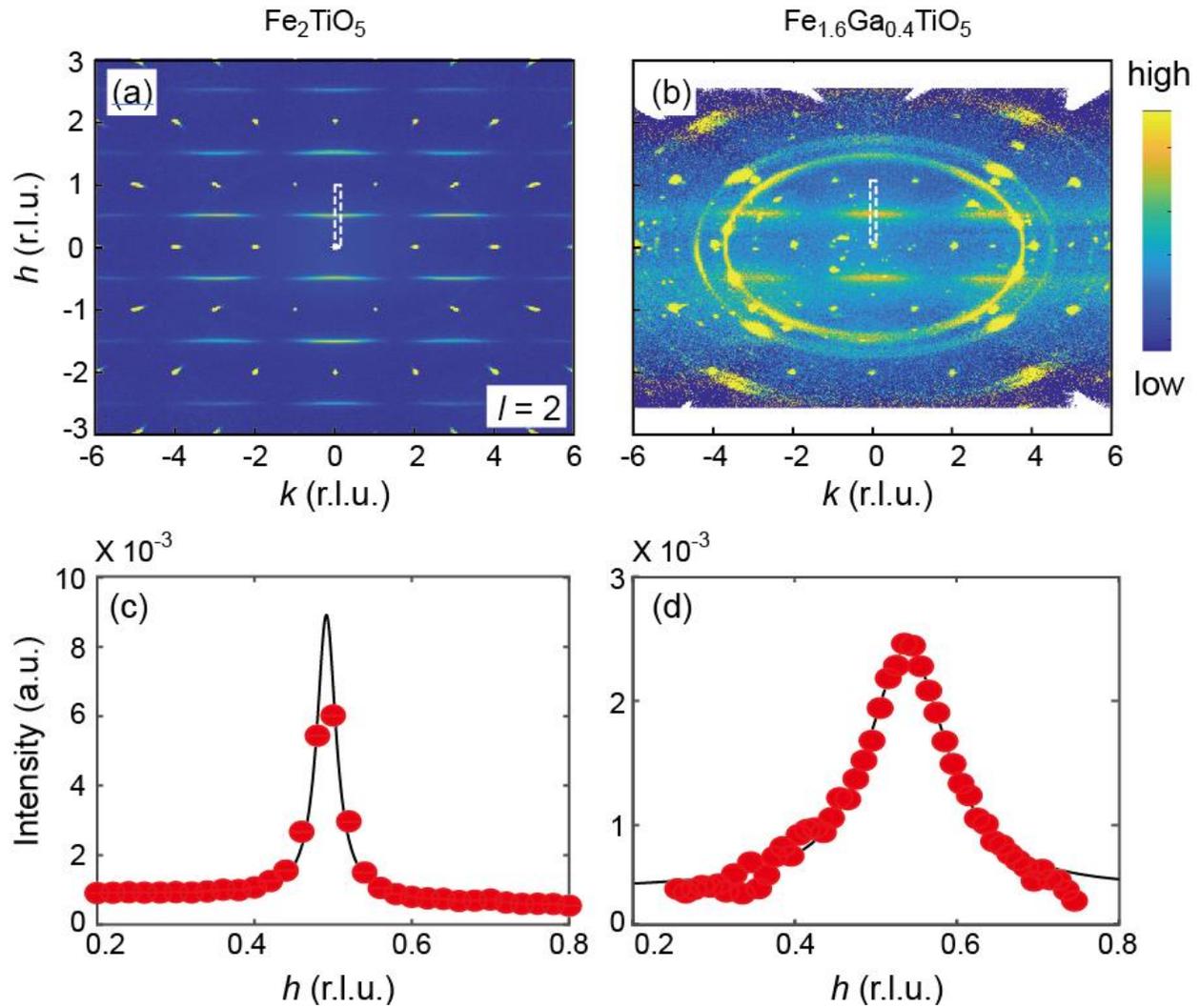

Figure 4.
Total neutron scattering intensity in the (*h*, *k*, 2) planes for $Fe_2TiO_5$ at T=5 K (a) and $Fe_{1.6}Ga_{0.4}TiO_5$ at T = 6 K (b). (c),(d) Total neutron scattering intensity along the h -direction as indicated by the dashed rectangle in (a) and (b), respectively. The intensity was fitted with single Lorentzian functions convoluted with a Gaussian resolution function. The correlation length along the *h* -direction is estimated to be 10.8±0.7 Å at 5 K and 7.5± 0.6 Å at 50 K for $Fe_{1.6}Ga_{0.4}TiO_5$ and 405±1.0 Å at 5 K and 32.1±0.9 Å at 60 K for $Fe_2TiO_5$

REFERENCES

[1] U. Atzmony, E. Gurewitz, M. Melamud, H. Pinto, H. Shaked, G. Gorodetsky, E. Hermon, R. M. Hornreich, S. Shtrikman, and B. Wanklyn, *Anisotropic spin-glass behavior in $Fe_2TiO_5$*, Physical Review Letters **43**, 782 (1979).

[2] P. G. LaBarre, D. Phelan, Y. Xin, F. Ye, T. Besera, T. Siegrist, S. V. Syzranov, S. Rosenkranz, and A. P. Ramirez, *Fluctuation-induced interactions and the spin-glass transition in $Fe_2TiO_5$*, Physical Review B **103**, L220404 (2021).

[3] A. P. Ramirez, *Thermodynamic measurements on geometrically frustrated magnets*, Journal of Applied Physics **70**, 5952 (1991).

[4] S. V. Syzranov, and A. P. Ramirez *Eminuscent phase in frustrated magnets: a challenge to quantum spin liquids*, Nature Communications 13, 2993 (2022).

[5] Note:, The crystals used for this study were obtained from the Bell Labs Crystal Archive, the contents of which can be found at the NSF sponsored website: http://www.crystalsamplearchive.org/, where further details may be found. The particular crystals used in this study were characterized using x-ray diffraction, and electron energy-loss spectroscopy, details of which are found in the Supplemental Material.

[6] F. Ye, Y. H. Liu, R. Whitfield, R. Osborn, and S. Rosenkranz, *Implementation of cross correlation for energy discrimination on the time-of-flight spectrometer CORELLI*, Journal of Applied Crystallography **51**, 315 (2018).

[7] S. Nagata, P. H. Keesom, and H. R. Harrison, *Low dc-field susceptibility of CuMn spin glass*, Physical Review B **19**, 1633 (1979).

[8] V. Cannella, and J. A. Mydosh, *Magnetic ordering in gold-iron alloys*, Physical Review B-Solid State **6**, 4220 (1972).

[9] U. Larsen, *Characteristic temperatures in the spin-glass AuFe*, Physical Review B **18**, 5014 (1978).

[10] H. Maletta, and W. Felsch, *Insulating spin-glass system $Eu_xSr_{1-x}S$*, Physical Review B **20**, 1245 (1979).

[11] A. P. Ramirez, *Strongly Geometrically Frustrated Magnets*, Annual Review of Materials Science **24**, 453 (1994).

[12] S. Nakatsuji, Y. Nambu, H. Tonomura, O. Sakai, S. Jonas, C. Broholm, H. Tsunetsugu, Y. M. Qiu, and Y. Maeno, *Spin disorder on a triangular lattice*, Science **309**, 1697 (2005).

[13] R. Fichtl, V. Tsurkan, P. Lunkenheimer, J. Hemberger, V. Fritsch, H. A. K. von Nidda, E. W. Scheidt, and A. Loidl, *Orbital freezing and orbital glass state in FeCr2S4*, Physical Review Letters **94**, 027601, (2005).

[14] H. J. Silverstein, K. Fritsch, F. Flicker, A. M. Hallas, J. S. Gardner, Y. Qiu, G. Ehlers, A. T. Savici, Z. Yamani, K. A. Ross, B. D. Gaulin, M. J. P. Gingras, J. A. M. Paddison, K. Foyevtsova, R. Valenti, F. Hawthorne, C. R. Wiebe, and H. D. Zhou, *Liquidlike correlations in single-crystalline Y2Mo2O7: An unconventional spin glass*, Physical Review B **89**, 054433, (2014).

[15] S. J. Garratt, and J. T. Chalker, *Goldstone modes in the emergent gauge fields of a frustrated magnet*, Physical Review B **101**, 024413, (2020).

[16] P. W. Anderson, B. I. Halperin, and C. M. Varma, *Anomalous low-temperature thermal properties of glasses and spin glasses*, Philosophical Magazine **25**, 1 (1972).

[17] M. Hermele, M. P. A. Fisher, and L. Balents, *Pyrochlore photons: The U(1) spin liquid in a S=1/2 three-dimensional frustrated magnet*, Physical Review B **69**, 064404 (2004).
[18] L. Savary, and L. Balents, *Coulombic Quantum Liquids in Spin-1/2 Pyrochlores*, Phys Rev Lett **108**, 037202 (2012).
[19] L. Savary, and L. Balents, *Quantum spin liquids: a review*, Rep Prog Phys **80**, 016502 (2016).
[20] S. D. Pace, S. C. Morampudi, R. Moessner, and C. R. Laumann, *Emergent Fine Structure Constant of Quantum Spin Ice Is Large*, Phys Rev Lett **127**, 117205 (2021).
[21] O. Benton, O. Sikora, and N. Shannon, *Seeing the light: Experimental signatures of emergent electromagnetism in a quantum spin ice*, Physical Review B **86**, 075154 (2012).
[22] S. V. Syzranov, O. M. Yevtushenko, and K. B. Efetov, *Fermionic and bosonic ac conductivities at strong disorder*, Physical Review B **86**, 241102 (2012).
[23] S. V. Isakov, K. Gregor, R. Moessner, and S. L. Sondhi, *Dipolar Spin Correlations in Classical Pyrochlore Magnets*, Phys Rev Lett **93**, 167204 (2004).
[24] C. L. Henley, *Power-law spin correlations in pyrochlore antiferromagnets*, Physical Review B **71**, 014424 (2005).
[25] C. L. Henley, *The "Coulomb Phase" in Frustrated Systems*, Annual Review of Condensed Matter Physics **1**, 179 (2010).
[26] A. Sen, K. Damle, and R. Moessner, *Fractional Spin Textures in the Frustrated Magnet $SrCr_{9p}Ga_{12-9p}O_{19}$*, Physical Review Letters **106**, 127203, (2011).
[27] A. D. LaForge, S. H. Pulido, R. J. Cava, B. C. Chan, and A. P. Ramirez, *Quasispin Glass in a Geometrically Frustrated Magnet*, Physical Review Letters **110**, 017203, (2013).